# Nanosized precipitates in H13 tool steel

# low temperature plasma nitriding

Zagonel et al.

# Nanosized precipitates in H13 tool steel low temperature plasma nitriding

L. F. Zagonel[a,b*], J. Bettini[a], R. L. O. Basso[b], P. Paredez[b], H. Pinto[d], C. M. Lepienski[e], F. Alvarez[b]

[a] Laboratório Nacional de Nanotecnologia, Centro Nacional de Pesquisa em Energia e Materiais, 13083-970, Campinas, Brazil

[b] Instituto de Física "Gleb Wataghin", Universidade Estadual de Campinas, Unicamp, P.O. Box 6165 Campinas, SP, 13083-970, Brazil

[d] Escola de Engenharia de São Carlos, Universidade de São Paulo, 13566-590, São Carlos/SP, Brazil

[e] Departamento de Física, Universidade Federal do Paraná, Brazil

Abstract

A comprehensive study of pulsed nitriding in AISI H13 tool steel at low temperature (400°C) is reported for several durations. X-ray diffraction results reveal that a nitrogen enriched compound ($\varepsilon$-$Fe_{2-3}N$, iron nitride) builds up on the surface within the first process hour despite the low process temperature. Beneath the surface, X-ray Wavelength Dispersive Spectroscopy (WDS) in a Scanning Electron Microscope (SEM) indicates relatively higher nitrogen concentrations (up to 12 at.%) within the diffusion layer while microscopic nitrides are not formed and existing carbides are not dissolved. Moreover, in the diffusion layer, nitrogen is found to be dispersed in the matrix and forming nanosized precipitates. The small coherent precipitates are observed by High-Resolution Transmission Electron Microscopy (HR-TEM) while the presence of nitrogen is confirmed by electron energy loss spectroscopy (EELS). Hardness tests show that the material hardness increases linearly with the nitrogen concentration, reaching up to 14.5 GPa in the surface while the Young Modulus remains essentially unaffected. Indeed, the original steel microstructure is well preserved even in the nitrogen diffusion layer. Nitrogen profiles show a case depth of about ~43 μm after nine hours of nitriding process. These results indicate that pulsed plasma nitriding is highly efficient even at such low temperatures and that at this process temperature it is possible to form thick and hard nitrided layers with satisfactory mechanical properties. This process can be particularly interesting to enhance the surface hardness of tool steels without exposing the workpiece to high temperatures and altering its bulk microstructure.

# 1. Introduction

The surface hardness, corrosion resistance, and wear resistance of steels can be improved by nitriding processes. [1,2,3,4,65] These thermo-chemical surface treatments promote the inward diffusion of nitrogen and its subsequent reaction with the alloying elements. [5,6,7] Depending on the nitriding temperature and nitrogen chemical potential at the material surface, the process may cause the formation of a continuous iron nitride surface layer. Underneath this layer, a diffusion zone will form containing nitrogen in solution and, in case of strong chemical affinity of nitrogen with the alloying elements, nitride compounds can be formed. In order to improve as much as possible the work piece surface properties a better understanding on the growing mechanism of nitrided surface layers is strongly justified.

Industrial nitriding processes are usually carried out by gaseous, plasma-assisted, or salt bath techniques. Among these, plasma nitriding is particularly attractive because of its relative high efficiency even at low temperatures. This is relevant to materials, such as stainless steels or tool steels, for which this condition is mandatory in order to minimize microstructural alterations and, consequently, maintain the corrosion resistance in the bulk material. [8,9,10,11] Moreovera review of relevant literature shows that a precise control of the surface characteristics and composition during plasma nitriding is a challenge, owing to the complexity of this non-equilibrium process. The proper setting of the process parameters (gas composition, gas pressure, ion current density, voltage, and temperature) and surface state (roughness, crystalline orientation, oxidation level, etc) are therefore indispensable to assure reproducibility of the process results. Furthermore, the bombardment of the work piece by ionized hydrogen and nitrogen species present in the plasma allows for an additional surface cleaning by physical and chemical sputtering ("etching") of the oxide layers usually present on the treated work pieces, thus adding one more variable to this complex process.[12,13]

As noted above, underneath the outmost nitride layer, nitrogen diffuses and (may) react(s) with alloying elements. These phenomena have been the focus of numerous studies attempting to point out key parameters affecting the whole process and the final material properties. [14,15] Recent studies reported by Mittemeijer and coworkers have shown the importance of the so called "excess" nitrogen on the diffusion process. [16,17,18,19] Such increased nitrogen concentrations represent the atomic fraction surpassing the quantity that should be expected in case that: (i) the alloying elements are fully consumed to form nitrides and (ii) the nitrogen equilibrium solubility in an unstrained iron matrix is achieved. The nitriding of AISI H13 hot work tool steel has also been intensively studied leading to enhanced surface hardness, thermal fatigue, and corrosion resistance. [5,6,20,21,22,23] In recent works, the mechanism of N diffusion at both low temperatures and short nitriding times (up to 5 hours) were reported by our group. Those studies were focused on the formation of the compound layer and of the diffusion zone as well as on the effect of the nitriding gas mixture ($N_2$ to $H_2$ ratio), nitriding temperature, and ion current density.[24,25,26,27] In those studies, we show that for the H13 steel, at 500°C, considerable decarburizing takes place at the surface including the dissolution of chromium carbides.[25] Even if no bulk hardness loss could be noticed after nitriding at this temperature, the observed microstructure changes may indicate that nitriding could have negative side effects at this temperature for this steel, particularly if the nitriding processes need to be repeated several times during the tool lifetime, as it is the case for some cutting tools.

In this paper we deal with a similar issue, i.e. the nitrogen diffusion mechanism at low temperature (400°C), but considering longer processing times (up to 36 hours). Since temperature does not affect significantly the plasma structure itself, this temperature regime maintains the effectiveness of plasma nitriding but modifies the kinetics of the processes inside the material, such as nitrogen diffusion as well as formation and dissolution of precipitates formed with the alloying elements. For instance, at 400°C, carbide dissolution in the diffusion zone is prevented.[26,27] Moreover, this temperature regime could be particularly interesting to avoid embrittlement, distortions or corrosion susceptibility which may result from excessive nitride precipitation at the higher temperatures normally used in the gaseous nitriding process.[28,29] The aim the present work is to determine the properties of thick (~50 μm) nitrided layers prepared at relatively low temperatures (~400°C) in the attempt to understand how precipitation, nitrogen diffusion, phase transformation and hardening takes place in this process regime. In this study, we show the effectiveness of low temperature plasma nitriding applied to H13 hot work tool steel with sufficient processing time to reach case depths with technological relevance.

## 2. Experimental procedures

The nitriding experiments were carried out using a commercial hot wall furnace plasma nitriding system operated with a pulsed current power supply working in a 50 μs pulse on and 150 μs pulse off.[30] A gaseous mixture of 20% $N_2$ + 80%$H_2$ was fed into the furnace controlled by mass flow meter controllers. A 400 Pa working pressure was used along the processing. As a compromise between plasma stability and current, the negative pulse was set to -380 V. The temperature was controlled by three independent heating elements and the working temperature was fixed at 400±1°C. The studied samples were nitrided under identical conditions but different treatment times, namely 1, 9, 16 and 36 hours.

The test samples were cut out from the interior of a single quenched and tempered AISI H13 steel block with bulk hardness of 7.5 GPa (quenched in air from 1025°C and tempered at 580°C for 2 h). The chemical composition of this steel is C, 2.5%; Mn, 0.4%; Si, 2.1%; Cr, 5.9%; Mo, 0.8%; V, 1%; Fe, balance (measured by inductive-coupled plasma-optic emission spectroscopy). The sample surfaces were mirror polished (final step was using 1 μm diamond paste) and cleaned in an acetone ultrasonic bath before loaded into the plasma nitriding furnace. The final surface roughness was about 40 nm ($R_a$), as measured by profilometry.

The phase composition of the nitrided layers was analyzed by X-ray diffraction (XRD) using monochromatic Cu Kα radiation and Bragg-Brentano (BB) θ-2θ geometry. The outmost surface zone was characterized using small angle XRD (referred to as SA geometry) at 3º incidence angle (effective penetration depth of ~0.1 μm). [31]

X-ray photoemission electron spectroscopy (XPS) was employed to gather information on the chemical state of nitrogen and alloying elements on the most outer material surface. The underneath layers were also scrutinized by XPS in samples previously *ex situ* grinded and polished (last step with 0.25μm diamond paste) and *in situ* cleaned with $Ar^+$ ion beam bombardment procedure. The XPS system is a VG-CLAM2 electron energy analyzer with a non-monochromatic Al Kα radiation source having a total resolution 0.85eV (X-ray line width plus analyzer).

The composition depth profiles were analyzed on mirror polished samples cross-sections by electron probe microanalysis (EPMA) employing a Cameca SX100 apparatus operated at 15kV using the Wavelength Dispersive Spectrometer (WDS). For cross-section preparation, the samples received an electrodeposited nickel protective layer after being cut, and were grinded and polished (last step with 0.25 µm diamond paste). The chemical composition was determined from the emitted characteristic X-ray intensities by their comparison with the corresponding intensities recorded from appropriated standards such as pure metals and compounds ($Fe_3C$, $Fe_4N$). The final element content was calculated applying the $\Phi(\theta z)$ method.[32] The estimated error from the EPMA measurements is lower than 0.5 at % for all reported results. The sample cross-sections were etched to reveal the material microstructure and observed by scanning electron microscopy (SEM) using a FEI Inspect F50. The etchant was a 5% Nital solution containing 0.1% concentrated hydrochloric acid. Transmission Electron Microscopy (TEM) analysis were performed in a JEOL 2100F microscope. Suitable thin samples were prepared applying the standard procedure, i.e., by gridding, polishing and dimpling down to 20 µm. Electron transparent areas were then obtained by low energy ion polishing.

Hardness tests were performed using a computer controlled nanoindenter equipped with a Berkovich diamond tip (Nanotech). The Oliver and Pharr method was used to calculate the hardness values from loading/unloading curves.[33] The precise position of each indentation was determined *a posteriori* by SEM. The maximum load was set to 400 mN with a dwell time of 15 s. Measurements were performed on polished cross-sections and on the surface of the samples. No piling up effect was considered in the analysis.

### 3. Results and Discussion

Figure 1 shows the XRD diffractograms obtained in the BB geometry for all the employed nitriding times. For comparison purposes, the result from SA XRD for the 1 h and 36 hour(s) nitrided samples are also displayed. The diffractogram recorded after 1h nitriding using the BB geometry reveals only the martensitic matrix phase, i.e., α-Fe. Similarly, the 9 hours nitrided sample shows a line pattern compatible with the matrix and, also, a tiny diffraction line at about 38° associated with the (110) planes of the ε-$Fe_{2-3}N$ phase.[34,35,36] The relative volume fraction of the ε-$Fe_{2-3}N$ phase increases significantly only after nitriding for 16 and 36 hs. On the other hand, SA XRD reveals that the ε-$Fe_{2-3}N$ phase is already formed during the first hour. Nevertheless, it exists only within the outmost sample surface and it is not observed when using the BB geometry. These results are in agreement with earlier observations showing that, upon plasma nitriding, the low nitrogen diffusion coefficient at 400°C favors the formation of nitrogen-rich iron nitrides, such as the ε-$Fe_{2-3}N$ phase. Indeed, the reactivity of the nitriding plasma (i.e., its efficiency) does not depend directly on the processes temperature. [25,37] Thus, owing to the high reactivity of the ion plasma process, the nitrogen surface concentration reaches a steady state stoichiometry characterized by the ε-$Fe_{2-3}N$ phase, during the first one hour of nitriding. [26,28]

Moreover, nitriding for 1 and 9 hours leads the diffraction lines associated to the α-Fe matrix to be shifted towards lower 2θ-angles, as shown in Fig. 1. For instance, the (110) reflection shifts from 44.87±0.05° in the untreated sample to 44.61±0.05° after 1h nitriding, whereas the (200) diffraction line shifts from 65.08±0.05° to 64.83±0.05° in the same cases. This shift is due to an increased lattice parameter of the α-Fe phase after nitriding. This increase in lattice parameter has two causes [39]: i) the nitrogen dissolved in the octahedral interstitial sites of the α-Fe lattice and adsorbed at nitride/matrix interfaces; ii) the positive volume misfit between the coherent nitrides and the martensitic matrix, which provokes a dilation of the matrix lattice and, consequently, an overall hydrostatic tensile stress in the α-Fe matrix. We note, however, that coherent nitride precipitates with few atomic layers thick are not detectable as separate diffraction lines by XRD measurements, since they diffract coherently with the surrounding martensitic matrix [43,44,45,54]. Thus, they only contribute to the broadening of the α-Fe or ε-Fe$_{2-3}$N diffraction lines.

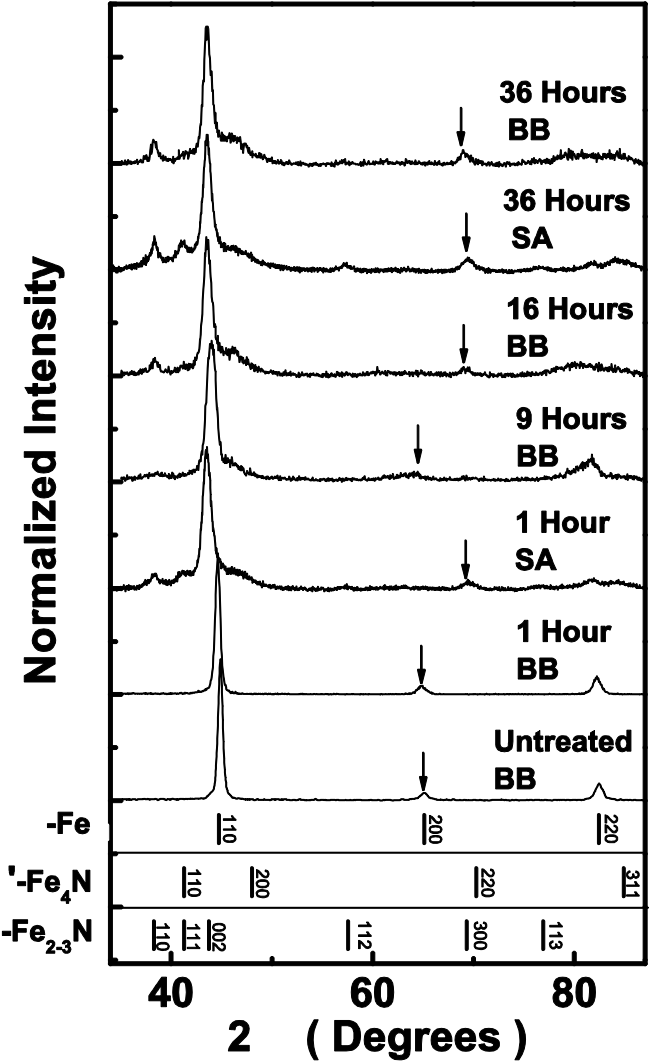

Figure 1: BB geometry X-ray diffractograms obtained in samples nitrided by 1, 9, 16 and 36 hours. SA results are shown for the sample nitrided for 1 and 36 hours. The arrow indicates the peak for (200) α-Fe or the (300) ε-Fe$_{2-3}$N for a reference of the most relevant phase on the surface.

As observed in the diffractograms, all the diffraction lines associated with the H13 matrix also become significantly broadened for longer nitriding times (see for instance the 220 reflection for the sample nitrided during 9 hours, Fig. 1). Nitrogen dissolved in the martensitic solid solution represents a first contribution to increasing line broadening. However, broadening purely caused by nitrogen dissolution within the α-Fe unit cell would cause a monotonous line width increase with the reflection order, i.e. $h^2+k^2+l^2$, without any dependence on particular crystallographic directions. [40] Nevertheless, the diffractograms show that the broadening of the α-Fe reflections is anisotropic with $\beta_{(200)} > \beta_{(211)} > \beta_{(110)}$, where β is the diffraction line breadth (see Table I). The same trend for the α-Fe line widths was encountered in [26]. The abnormal increase of line width associated with the (200) α-Fe planes reflections suggests once again that nanosized coherent particles have precipitated within the steel matrix. As discussed in the literature coherent plate-like nitrides develop at the initial nitriding stages in Fe-Cr-V alloys.[23,41,42,43,44] These nanosized platelets (cubic CrN, VN) growth follows the Bain orientation relationship with respect to the α-Fe matrix [42,54]:

$(100)_{(Cr,V)N} // (100)_{\alpha-Fe}$; $[110]_{(Cr,V)N} // [001]_{\alpha-Fe}$

Because of this growth orientation of the coherent precipitates, line broadening is more significant for the (h00)-type planes of the steel matrix, which are directly connected to the lattice of the nanosized precipitates and therefore mostly distorted. [23].

Tables

Table I: Full width at half maximum (FWHM) of the α-Fe diffraction lines.

| hkl | h²+k²+l² | Untreated material | After 1 hour treatment | After 9 hour treatment |
|-----|----------|--------------------|-----------------------|-----------------------|
| 110 | 2        | 0.473              | 0.508                 | 1.085                 |
| 200 | 4        | 0.903              | 0.901                 | 2.653                 |
| 211 | 6        | 0.791              | 0.828                 | 2.376                 |
| 220 | 8        | 1.036              | 1.07                  | 1.225                 |
| 310 | 10       | 1.884              | 1.907                 | 2.367                 |

The microstructure of the sample nitrided by 36 hours is revealed by SEM (Fig. 2 (a) and (b)). The cross section image displays the tempered martensitic microstructure of the H13 steel, where some globular precipitates, identified as vanadium carbides by EDX analyses, are embedded (Fig. 2(b)). The complexity of the microstructure with small grain size is also observed. Significant microstructural differences are not apparent by comparing the diffusion zone and the original matrix (Fig. 2 (a)). Similarly, the bulk region of this sample has the same microsctructure as untreated samples (not shown). As shown later on, the nitrogen concentration varies between ~16 at.% to ~3 at.% moving from the top to the bottom of the image without noticeable microstructure changes (Figure 4). In contrast to previous reports also involving low temperature (370ºC and 400°C) H13 steel nitriding experiments, increasing nitrogen concentration does not seem to occur along the grain boundaries of the diffusion zone. [25,26] This effect has also been observed in both ion- and gas nitrided materials. [46,47] In the case of strong preferential nitrogen diffusion through grain boundaries, it is observed a peak in the nitrogen profile when a grain boundary is crossed.[63] Also, these regions appear dark on SEM-BSE images. [25,27]

Moreover, the H13 tool steel investigated in the present paper exhibits a matrix free of chromium precipitates, i.e. most of the chromium content appears to remain dissolved within a α-Fe solid solution. This is in contrast to previous observations that the tempering applied to the H13 steel after quenching leads to an extensive precipitation of chromium-rich particles in the initial state prior to nitriding. [12,27,65] These observations suggest that the presence of chromium preferentially in the solid solution could be linked to the enhanced nitrogen bulk diffusion. This agrees with the predictions made by Krawitz [48] and Gouné [49] that the thermodynamic interaction between nitrogen and the alloying elements determines the extent of nitrogen up take in some iron-based materials. Thus, chromium, having high affinity with nitrogen, seems to induce nitrogen diffusion through the bulk material, increasing nitrogen trapping probability (i.e. adsorption) at dislocations and interfaces present in the tempered martensitic microstructure. This effect is also observed in powder metallurgy alloys containing vanadium. [50]

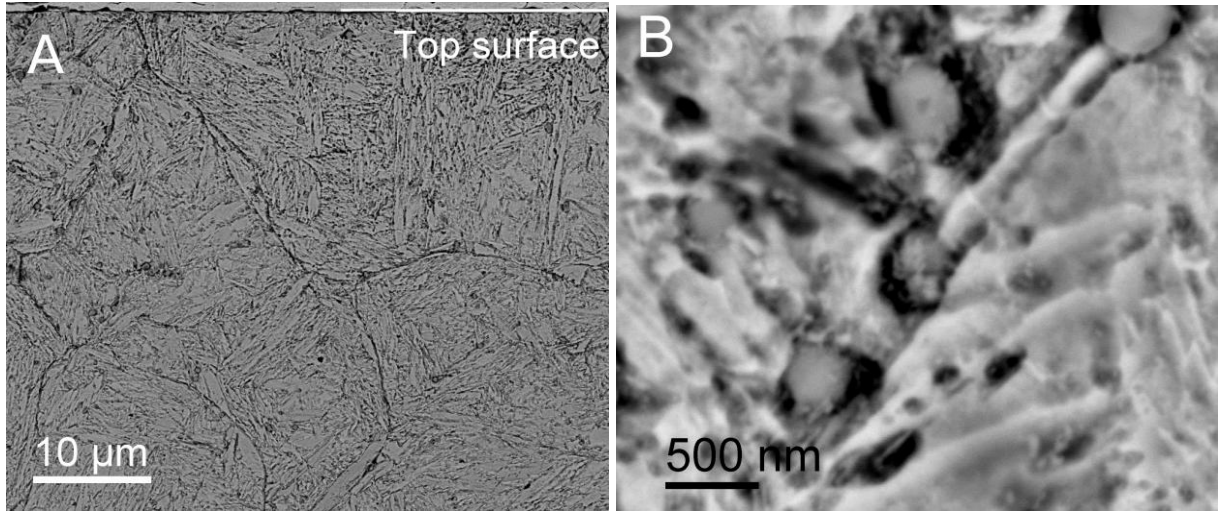

Figure 2: SEM micrograph (secondary electrons) of the sample nitrided by 36 hours. (a) Low magnification image showing the microstructure on the diffusion layer. The sample surface is on the top of the figure and discontinuous layer on top is a protective nickel coating for sample preparation purposes. (b) Detail on vanadium carbide precipitates.

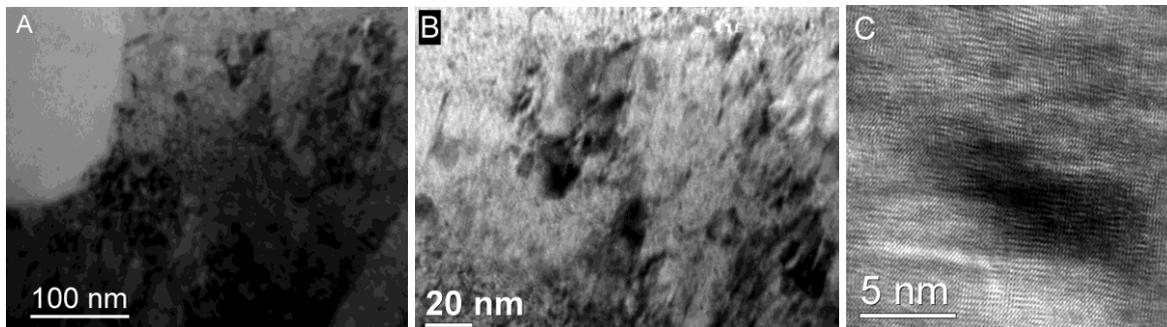

Figure 3: TEM images from the sample nitrided for 36 hours. (a) Bright field image showing a big feature on the top-left corner identified by EELS as a vanadium carbide. (b) Dark-gray areas are identified as nanoprecipitates of chromium nitride. (c) HR-TEM showing that nanosized precipitates can be coherent with the steel matrix.

To complete these observations, TEM analysis (including EELS) has been carried out, allowing the observation of fine nanometric precipitates. [45,54,62] Figure 3 shows a bright field TEM images from the sample nitride during 36 hours just a few microns below the sample surface. Although chemical mapping was not possible due to the magnetic character of the sample, it was possible to perform EELS analysis in selected areas by focusing the electron beam to a small spot (~5 nm). Therefore, as revealed by EELS, the feature observed on the top-left of the Fig. 3(a) is compatible with vanadium carbide. The dark regions with fine structures are identified as fine dispersed chromium nitride precipitates, with low carbon content. Fig. 3(b) shows a closer look on this area. Dark-gray areas show mainly nitrogen, chromium and iron (no quantification was attempted since the exact probed volume could not be determined accurately). Further EELS analysis indicates that light-gray areas in Fig. 3 (b) are similar in composition to dark-gray areas, indicating the presence of both chromium and nitrogen in the steel matrix possibly as very fine precipitates. Indeed, by a careful observation of Fig. 3(b), it is possible noting very fine features, about 1-2 nm in width, which could be ascribed to chromium nano precipitates. The coherent nature of some precipitates could be confirmed by HR-TEM images. As an example, Fig. 3(c) shows atomic columns on one bigger chromium nitride nanosize precipitate while in its vicinities, smaller precipitates are observed with faint contrast over the steel matrix. This precipitate has a projected area of about 4x6 nm$^2$. Finally, in agreement with SEM investigations, no micrometric nitride precipitate was found.

Figure 4(a) shows the elemental profile from the sample nitrided for 36 hours obtained by electron probe micro-analysis (EPMA) close to the surface. These profiles were taken as line scans perpendicular to the surface and contain fluctuation due to the local microstructure variations. Figure 4(b) shows a micrograph acquired just after the profile acquisition showing the location of the measured points. Vanadium carbides are identified in points where vanadium and carbon concentrations are far larger than those from the matrix average ones. At these locations, molybdenum concentration also increases slightly indicating that the precipitates contain small molybdenum quantities in addition to vanadium. The chromium depth profile is fairly homogeneously distributed, indicating that chromium carbides or nitrides do not form large precipitates, in agreement with TEM analysis. Peaks associated with vanadium and carbon concentrations along their depth profiles reveal that vanadium carbides are not dissolved in the nitrided layer, also confirmed by TEM results [27]. Similar features were observed in other samples prepared by nitriding for 9 and 16 hours treatments. At such low temperatures (400°C), the precipitation kinetics might be too slow to allow the formation of microscopic nitride precipitates or the dissolution of carbides present in the matrix. [51,52]

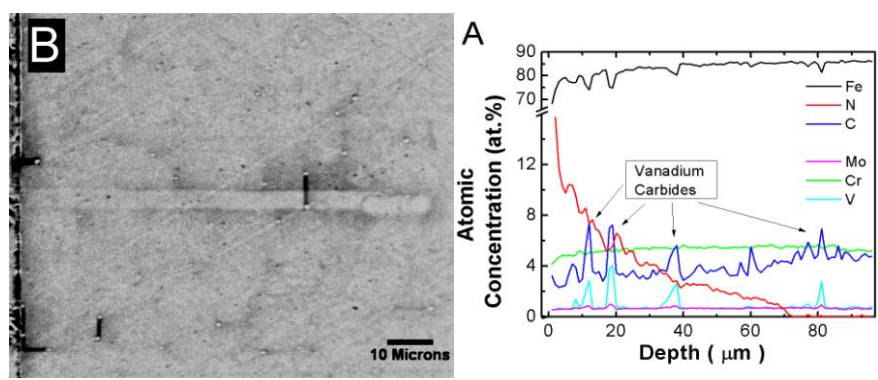

Figure 4: (a) Profile of principal elements from 36 hours nitrited sample. Concentrations of V and C above average are identified as Vanadium Carbides. (b) Micrograph taken just after the composition profile (WDS) acquisition showing the measured region and the electron probe line scan.

For comparison, the nitrogen profiles of samples nitrided for 9, 16, and 36 hours are shown in Figure 5. For these samples, the nitrogen surface concentration is close to 16 at%. These concentrations stem from the presence of (possibly dispersed) ε-Fe$_{2-3}$N iron nitrides in the material surface convoluted by the electron excitation volume probed, which is much larger than the iron nitride layer. Beneath of a ~3 μm layer (i.e. beneath iron nitrides), the nitrogen concentration measured on the cross-sections by EPMA is ~12%. This amount is still far larger than the nitrogen solubility limit (~0.1 at.% at 400°C) in un-strained iron matrix.[53] Moreover, this nitrogen concentration is also larger than the quantity that would be necessary to consume the full Cr and V contents (~ 7.0 at.%) in the H13 steel to form nitrides (moreover, V is observed to be forming carbides and hence unavailable to form nitrides). Also, the additional line broadening observed for all α-Fe diffraction features after nitriding times longer than 9 hours indicates that interstitial excess nitrogen also contributes to strain the steel matrix. [7,34] Therefore, nitrogen appears to be trapped at the interfaces of the nanosize coherent precipitates observed on the TEM.[54] Since the tempered martensitic microstructure of this steel is very complex, high densities of dislocations and interfaces could favor the presence of immobile excess nitrogen. [18] Indeed, nitrogen must be at interstitial sites, interfaces, and forming precipitates to explain such high nitrogen concentrations in the steel matrix. In this complex scenario, it is observed a coincidence in the nitrogen profiles (shown in the inset of Fig. 5.) if they are scaled by normalizing the diffusion depth by the square root of the nitriding time. This could indicate that the mobile nitrogen has diffused under the effect of traps.[55] Also, even for 9 hours nitriding at 400°C, the depth of the diffusion zone reaches more than 40 μm, indicating the efficiency of this process even at such low temperatures.[23]

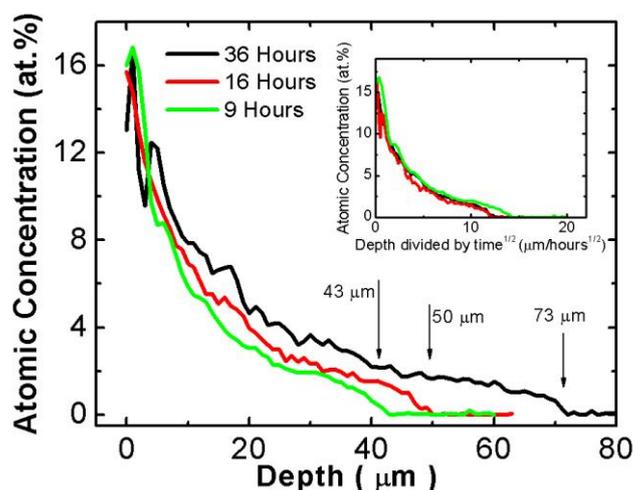

Figure 5: Nitrogen concentration profile for different studied nitriding times. The arrows indicate the nitrogen penetration depth for 9, 16 and 36 hours.

As expected, the presence of the ε-$Fe_{2-3}$N phase and coherent nanoprecipitates on the surface increases the surface hardness of the studied samples by nearly 100%, see Table II.[7,25] We attributed the observed slight surface hardness differences to the increasing thickness of the compound layer on time. Moreover, the surface Young's modulus is not significantly changed by the nitriding process, indicating that the modified surface maintain its original elasticity, see Table II. We note that in a bimetallic two dimension planar structure, the radius of curvature has a minimum for equal Young´s modulus of the materials forming the junction.[56] Therefore, it is a very interesting finding since the Young´s modulus matching between the hard nitride case and the H13 substrate diminishes the chances of delamination due to interface shear stress.

The hardness was measured on the cross-sections by nanoidentation as function of depth and the results are correlated to the nitrogen profiles (shown in Figure 6). Moreover, it is observed a fairly good linear correlation between hardness and nitrogen concentration within the error bars, as previously reported by other studies. [57,58,59] The correlation coefficient obtained from the statistical analysis is $R^2 = 0.91$ and the relation is $H - H_0 = 0.73\,[N]$ (GPa/at.%), where $H_0$ is the bulk hardness, $7.5\pm0.4$ GPa. This behavior is consistent with the occurrence of a homogeneous precipitation of nanosized nitrides.[60,18,54]

Table II: Surface (~1 μm depth) hardness and Young's modulus for different treatment times.

| Nitriding Time (hours) | Young's modulus (GPa) | Surface Hardness (GPa) |
|---|---|---|
| 0 (matrix) | 260 ± 7 | 7.5 ± 0.4 |
| 9 | 250 ± 8 | 14.0 ± 0.7 |
| 16 | 250 ± 12 | 14.3 ± 0.9 |
| 36 | 250 ± 20 | 14.4 ± 0.7 |

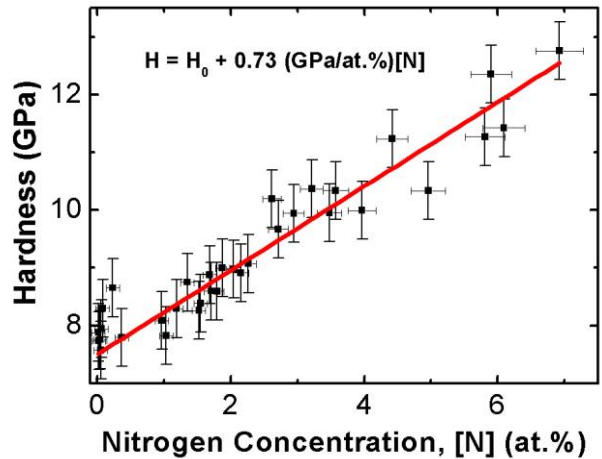

Figure 6: Hardness variations H vs. nitrogen concentration [N]. The best straight line fitting is indicated in the plot.

XPS measurements can provide information on the average local chemical bonding state of the alloying elements. Figures 7 displays the spectra associated with the N1s and $Cr_{3/2}$ electrons core levels obtained at different depths from the sample nitrided by 16 hours. As observed, for these elements the binding energy and band shape are similar for all mapped depths. Moreover, the Cr binding energy corresponds fairly well to the value expected for the solid solution or nanosized precipitates of the element in iron α-Fe matrix.[61] Regarding nitrogen, the peaks are relatively enlarged, phenomenon steaming from nitrogen in different chemical states, i.e. dissolved in solid solution, adsorbed at interfaces, and forming different fine precipitates. Indeed, previous in situ XPS experiments, showed similar results, confirming that at such low temperatures, no clear chemical state is reached for nitrogen. [26,27,64] In the assumption of extensive formation of nitrides, binding energy shifts are expected, as previously reported in experiments performed at higher temperatures [26,27,57]. We remark that the residual nitrogen concentration at a depth of 60 μm is most likely due to the incertitude in the depth measurement.

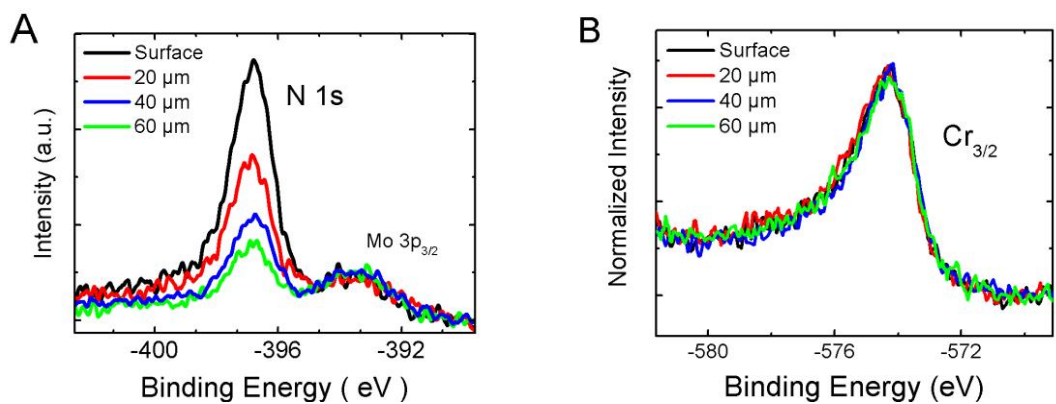

Figure 7: XPS spectra associated with the N1s and Crd$_{3/2}$ core level electrons. The curves show no change in the chemical state of the elements as function of depth (16 hr sample).

### 4. Conclusions

The effect of nitriding AISI H13 tool steel at a low temperature (400°C) is reported. The high plasma reactivity even at this temperature favored the formation of the ε-Fe$_{2-3}$N phase at the barely surface of the material. High nitrogen concentration was observed in the diffusion layer of the studied samples, indicating the presence of excess nitrogen (concentration above expected values). Nitrogen diffuses forming coherent nanosized precipitates while original carbides present in the matrix are not dissolved during the process. Indeed, large sized precipitation was inhibited by the low temperature process even for high nitrogen concentrations (above 12 at% underneath the surface) and, consequently, nitrogen diffuses as deep as 40 to 70 μm in the material bulk. These findings can be explained by the excess nitrogen in the matrix interstitial sites, the α-Fe matrix expansion due to coherent nanoprecipitates, and finally immobile excess nitrogen trapped at the interfaces of the complex steel structure. This scenario is supported by the linear correlation of hardness and nitrogen concentration that suggests progressive homogeneous nanosize precipitation (contrary to coarse precipitation leading to hardness decrease). Moreover, the similar chemical bonding state of nitrogen and alloying elements suggested by the XPS measurements obtained at several depths of the treated sample support this interpretation. Finally, these results support the idea that plasma nitriding processes performed at lower temperatures is effective to form nanosized precipitates, thus enhancing surface hardness without affecting bulk microstructure or hardness. Therefore, this process could be indicated for tool steel applications demanding increased surface hardness for which high nitriding temperature processes are not acceptable due to workpiece distortion or change in bulk mechanical properties.

**Acknowledgments**

Part of this work was performed during a stay of LFZ in the Prof. E. J. Mittemeijer group at the Max Planck Institute (MPI) for Metals Research in Stuttgart, Germany. The authors are grateful to Mrs S. Haug from the MPI for assistance with the electron probe microanalysis measurements. This work was partially supported by FAPESP, grant # 05/53926-1. FA and FLZ are CNPq fellows. HP thanks the financial support of the DFG project 444Bra-113/25/0-1